\documentclass[twocolumn]{article}

\usepackage[left=1in,right=1in,top=1in,bottom=1in]{geometry}
\usepackage{tikz}
\usetikzlibrary{calc,patterns,angles,quotes}

\usepackage{algorithm}
\usepackage{algpseudocode}

\usepackage{amsmath}
\DeclareMathOperator{\EX}{\mathbb{E}}
\usepackage{amsfonts}
\usepackage{authblk}
\usepackage{url}
\usepackage{graphicx}
\usepackage{subcaption} % Add this line for subfigures
\usepackage{hyperref}
\usepackage[explicit]{titlesec}
\renewcommand{\thesection}{\Roman{section}}
\renewcommand{\thesubsection}{\thesection.\Alph{subsection}}
\renewcommand{\thesubsubsection}{\thesubsection.\arabic{subsubsection}}

\titleformat{\section}{\large\scshape\centering}{\thesection.\space}{0pt}{#1}[]
\titlespacing*{\section}{0pt}{0.5\baselineskip}{0pt}
\titleformat{\subsection}{\normalsize\itshape}{\Alph{subsection}.\space}{0pt}{#1}[]
\titlespacing*{\subsection}{0pt}{0.5\baselineskip}{0pt}
\titleformat{\subsubsection}{\normalsize\itshape}{\arabic{subsubsection}.\space}{0pt}{#1}[]
\titlespacing*{\subsubsection}{0pt}{0.5\baselineskip}{0pt}
\usepackage{geometry}
\newgeometry{left=2cm, right=2cm, top=2.5cm,bottom=3.5cm}
\makeatletter
\renewcommand{\fnum@figure}{Fig. \thefigure}
\renewcommand{\fnum@table}{Tab. \thetable}
\makeatother

\usepackage{nopageno}
\renewcommand{\abstractname}{}
\title{\textbf{\normalsize Stochastic Continuation of Trajectories in the Circular Restricted Three-Body Problem via Differential Algebra}}
\author[(1), (2)]{Giacomo Acciarini}
\author[(2)]{Nicola Baresi}
\author[(3)]{David J. B. Lloyd}
\author[(1)]{Dario Izzo}
\affil[(1)]{Advanced Concepts Team, European Space Agency, European Space Research and TechnologyCentre (ESTEC), Keplerlaan 1, 2201 AZ Noordwijk, The Netherlands}
\affil[(2)]{Surrey Space Centre, University of Surrey, GU2 7XH, Guildford, United Kingdom}
\affil[(3)]{Department of Mathematics, University of Surrey, GU2 7XH, Guildford, United Kingdom}

\date{}  % don't display a date
\begin{document}
\let\oldhat\hat
\renewcommand{\vec}[1]{\pmb{\mathrm{#1}}}
\renewcommand{\hat}[1]{\oldhat{\pmb{\mathrm{#1}}}}
\newcommand\numberthis{\addtocounter{equation}
{1}\tag{\theequation}}

\maketitle

\begin{abstract}
\vspace{-1.3\baselineskip}
\textbf{\emph{\quad Abstract} -
Numerical continuation techniques are powerful tools that have been extensively used to identify particular solutions of nonlinear dynamical systems and enable trajectory design in chaotic astrodynamics problems such as the Circular Restricted Three-Body Problem. However, the applicability of equilibrium points and periodic orbits may be questionable in real-world applications where the uncertainties of the initial conditions of the spacecraft and dynamical parameters of the problem (e.g., mass ratio parameter) are taken into consideration. Usually, the robustness of a candidate trajectory is tested via a two-step approach, whereby trajectories
are first designed in a deterministic scenario, and then Monte Carlo methods are a posteriori used to check their robustness. While this strategy is ubiquitous in preliminary mission design, it can however lead to time-consuming and potentially not robust solutions, meaning that the found trajectories are not designed to account for uncertainties. Instead, the robustness of the deterministic optimal solutions is usually ensured.
Due to uncertain parameters and initial conditions, the spacecraft might not follow the reference periodic orbit owing to growing uncertainties that cause the satellite to deviate from its nominal path. Hence, it is crucial to keep track of the probability of finding the spacecraft in a given region. Building on previous work, we extend numerical continuation to moments of the distribution (i.e., stochastic continuation) by directly continuing moments of the probability density function of the spacecraft state. Only assuming normality of the initial conditions, and leveraging moment-generating functions, Isserlis' theorem, and the algebra of truncated polynomials, we propagate the distribution of the spacecraft state at consecutive surface of section crossings while retaining a symbolic map of the final moments of the distribution that depend on the initial mean and covariance matrix only. While the technique is only valid for initial Gaussian distributions, it does not assume that the distribution maintains its normality throughout the integration. The symbolic Poincaré map can then be directly used to evaluate the final moments of an initial distribution, as a function of the initial mean and covariance. This can therefore be used to accelerate the evolving step in the stochastic continuation procedure.
The goal of the work is to offer a differential algebra-based general framework to continue 3D periodic orbits in the presence of uncertain dynamical systems. The proposed approach is compared against traditional Monte Carlo simulations to validate the uncertainty propagation approach and demonstrate the advantages of the proposed in terms of uncertainty propagation computational burden and access to higher-dimensional problems.}
\end{abstract}

\section{Introduction}

Mission design in the presence of uncertainties is usually tackled via a two-step approach. First, trajectories are designed in deterministic settings, and then, perturbations are applied based on the given uncertainties, which can come from either incomplete knowledge or random errors~\cite{feng2019survey}. While this procedure is popular in preliminary mission analysis, it can however be computationally cumbersome (due to the many trial-and-error procedures needed before converging to a given design), and it is not guaranteed to find solutions that fulfill scientific objectives while complying with mission requirements. In particular, when designing missions to complex dynamical systems, such as three-body environments, the situation can be particularly challenging, due to navigation errors and environment uncertainties~\cite{scheeres2016orbital}. An example is the case of missions towards small Solar System bodies, whose geophysical properties are generally poorly known a-priori, and the navigation errors can reach substantial values relative to the orbits that have to be flown~\cite{fodde2023robust}. 

A central role in the exploration of three-body systems is played by periodic orbits: being able to design robust periodic orbits would be essential to improve and enhance the exploration of our Solar System~\cite{szebehely2012theory}. 
Following recent work from the authors~\cite{acciarini2023stochastic}, we propose a stochastic continuation approach to the problem of designing spacecraft trajectories within the framework of the Earth-Moon circular restricted three-body problem with uncertainties in both the spacecraft initial conditions and mass ratio parameter of the primaries. By investigating moment maps~\cite{barkley2006moment} in cases where initial conditions and mass ratio parameter are uncertain, we look for solutions that have bounded properties, while also maintaining the periodicity condition, on average. The contribution of this work consists of extending the previous setup to three-dimensional orbits, as well as introducing a differential algebra (DA)-based approximation to alleviate the propagation of the moments of the distribution on Poincaré maps, compared to the Monte Carlo method.

\section{Background}
\subsection{Circular Restricted Three-Body Problem}
The  CR3BP is defined as the problem of three-body bodies when the mass of the third body is negligible (and therefore considered massless) with respect to the two primaries, and when the two primaries move in a circular orbit about the system's barycenter (e.g. the motion of a spacecraft in the Earth-Moon system). Writing the equations of motion in a synodic reference system that is centered in the barycenter and co-rotates with the circular motion of these about their mutual barycenter, the motion of the third body can be fully described by the following set of 2nd order ordinary differential equations (ODEs)~\cite{koon2000dynamical}:
\begin{align}
    \begin{split}
    \ddot{x}-2\dot{y}&=-\dfrac{\partial \bar{U}}{\partial x}\\
    \ddot{y}+2\dot{x}&=-\dfrac{\partial \bar{U}}{\partial y}\\
    \ddot{z}&=-\dfrac{\partial \bar{U}}{\partial z}
    \end{split}
    \label{eq:equations_of_motion}
\text{,}
\end{align}

where $\mu=m_2/(m_1+m_2)$, with $m_1>m_2$, is the mass ratio parameter, $\bar{U}=-1/2(x^2+y^2)-(1-\mu)/r_1-\mu/r_2$ is the effective potential, $r_1^2=(x+\mu)^2+y^2+z^2$, and $r_2^2=(x+\mu-1)^2+y^2+z^2$.

\subsection{Periodic Orbits in the CR3BP}
\label{sec:periodic_orbits_in_the_cr3bp}

If a periodic orbit exists, with period $T$, it satisfies $\vec{F}(\vec{x}_0)=0$, where:

\begin{equation}
    \vec{F}(\vec{x}_0)=\phi(T,t_0;\vec{x}_0)-\vec{x}_0=0
    \text{,}
    \label{periodicity_condition}
\end{equation}
with $\phi$ being the flow of the dynamical system described by Eq.~\eqref{eq:equations_of_motion}.
These equations, together with Eq.~\eqref{eq:equations_of_motion}, define a boundary value problem (BVP), whose solutions can be found via a Newton procedure~\cite{baresi2017spacecraft}. Starting with an initial guess for the vector, which we refer to as $\vec{x}_{k,0}$, we can write its Taylor expansion as:

\begin{equation}
    \vec{F}(\vec{x}_{k+1,0})=\vec{F}(\vec{x}_{k,0})+DF(\vec{x}_{k,0})(\vec{x}_{k+1,0}-\vec{x}_{k,0})+HOT
    \text{,}
\label{eq:Taylor_expansion_to_derive_newton_method}
\end{equation}
where $DF$ is the Jacobian of $\vec{F}$ w.r.t. the initial conditions and $HOT$ indicate the higher order terms. From this, we can directly define an iterative scheme to find the next iterate, as a function of the previous iterate:
\begin{equation}
    \vec{x}_{k+1,0}=\vec{x}_{k,0}-DF^{-1}(\vec{x}_{k,0})\vec{F}(\vec{x}_{k,0})
    \text{.}
    \label{eq:newton_method}
\end{equation}
This can lead to convergence to a periodic orbit if the Jacobian is invertible and the initial guess is close enough. However, in the CR3BP, all the points along the periodic orbit still satisfy the periodicity condition, and the system is also a Hamiltonian autonomous system, making the periodic orbits organized in one parameter families~\cite{Meyer2017}.

These features introduce degeneracies in the BVP, which make the problem ill-posed and not invertible. The phase condition and pseudoarclength continuation are generally used to overcome these singularities~\cite{seydel2009practical}. 
Alternatively, periodic orbits can also be found in the reduced system of four-dimensional coordinates, using Poincaré maps~\cite{poincare1893methodes}. These discrete maps are defined as the intersection of a periodic orbit of a continuous dynamical system, with a lower-dimensional subspace (also referred to as Poincaré section). In this setting, periodic orbits are seen as fixed points of the Poincaré map. Similarly to before, this procedure implies starting from an initial guess and setup a Newton procedure~\cite{doedel1991numerical1, doedel1991numerical2, howell1987numerical}. This is possible in the deterministic system because we have explicit access to the Jacobian of the vector-valued function. However, this is generally not the case for stochastic systems, since the Jacobian of the vector-valued function might not be easily accessible, or analytically computable at all. In Sec.~\ref{sec:stochastic_continuation} and \ref{sec:periodic_orbits_in_the_cr3bp}, we, therefore, introduce stochastic continuation and discuss our proposed setup to find and continue periodic orbits in the CR3BP, in the presence of uncertainties.
\subsection{Stochastic Continuation}
\label{sec:stochastic_continuation}
Robust mission design is related to the necessity of planning missions in uncertain environments, with noisy observations~\cite{surovik2017reactive, fodde2023robust, chappaz2015trajectory}. In the last years, the problem of continuing particular solutions of dynamical systems in the presence of uncertainties has been investigated, and a framework to continue these systems has been developed and applied for the study of equilibria in several Hamiltonian systems~\cite{barkley2006moment, willers2020adaptive}. Building on these, we have extended the techniques for the study of periodic orbits in the planar CR3BP~\cite{acciarini2023stochastic}. While the above work was instrumental in building the framework for stochastic continuation of periodic orbits, only the planar case was studied, and the considered uncertainties were still low dimensional. In this work, the objective is to extend the formulation to the CR3BP, and better investigate if the computational load of mapping uncertainties on Poincaré maps can be reduced by means of the differential algebra of Taylor polynomials (w.r.t. Monte Carlo sampling). These aspects will be discussed in Sec.~\ref{sec:periodic_orbits_in_the_stochastic_system} and \ref{sec:da_based_moments_evolution_on_poincare_maps}.

\section{Method}
\label{sec:method}
\subsection{Periodic Orbits in the Stochastic System}
\label{sec:periodic_orbits_in_the_stochastic_system}
We assume that both initial conditions and mass ratio parameter are random variables, which can be sampled from the probability density functions (pdf) describing their behavior, i.e., $\mu \sim p(\mu)$ and $\vec{x}_0\sim p(\vec{x}_0)$, where $p$ is the probability density function. Our objective is to study the $k$-th moment of the pdf that characterizes the state at future crossings of a user-defined surface of section (SoS). In particular, since we are interested in periodic orbits, we seek to find solutions that stay, on average, around the same nominal trajectory, while also maintaining a certain ``boundness" around the nominal solution. Following an approach similar to the deterministic case discussed in Sec.~\ref{sec:periodic_orbits_in_the_cr3bp}, we seek these solutions on a Poincarè mapping. Hence, we confine the study to a reduced state $\vec{y}$ that describes the state of the system on a Poincarè map. 
The goal is, given a certain uncertain environment (i.e., a certain pdf for the mass ratio parameter) to find the required initial pdf for the reduced state, which allows the spacecraft to maintain an average periodic orbit, while also ensuring that the directions of maximum stretch of its covariance are bounded within user-defined thresholds. 

Since we are studying a dynamical system described by the equations of motion shown in Eq.~\eqref{eq:equations_of_motion} the description of the evolution of the probability density function at future times is described by the Fokker-Planck equation (in this case with zero diffusion term)~\cite{gardiner1985handbook}. The more general problem of finding periodic solutions to the Fokker-Planck equation requires the matching of all the moments of the distribution. This is a broader problem than the one we focus on and it was discussed in several publications~\cite{ji2019existence, jiang2020existence, ji2021convergence}. In these works, it is shown how it can be very hard to even prove the uniqueness and/or existence of solutions, for general dynamical systems, without additional properties or constraints. In the case where no diffusion is present, a necessary condition for periodic solutions to exist is that bounded solutions exist~\cite{massera1950existence}.

Given these difficulties and the importance of bounded solutions, we propose a more confined setup, where we limit the study to short time horizons (i.e., one period), and only to the first two moments of the distribution. 
Moreover, we formulate the stochastic continuation problem as follows:

\begin{equation}
\begin{cases}
&\textrm{given} \ \Sigma_{\vec{y}_f} \\
&\textrm{find} \ \bar{\vec{y}}_f\\
&\textrm{such that}: \ \vec{G}(\bar{\vec{y}}_f; \Sigma_{\vec{y}_f})=\vec{0}\\
\end{cases}
    \text{,}
    \label{eq:vector_valued_function_stochastic_second_moment}
\end{equation}
with:
\begin{equation}
    \vec{G}(\bar{\vec{y}}_f;\Sigma_{\vec{y}_f})
    =\EX[P^{-1}(\pmb{y}_f;\mu)]-\begin{bmatrix}
    \bar{x}_f\\
    \bar{z}_f\\
    \bar{v}_{x,f}\\
    \bar{v}_{z,f}\\
    \bar{C}
    \end{bmatrix}
    \label{eq:G_formulation_stochastic_cont}
\end{equation}
where $P^{-1}$ is the inverse of the Poincaré map that describes how initial states are mapped onto the surface of section, and $\EX$ is the expectation operator (e.g., $\EX[x]=\int_{-\infty}^{\infty}x p(x)dx$, with $p$ probability density function of $x$). Expanding the elements of Eq.~\eqref{eq:G_formulation_stochastic_cont}, we define:
\begin{equation}
    \vec{G}(\bar{\vec{y}}_f;\Sigma_{\vec{y}_f})
    =\begin{cases}
    \EX_{\vec{y}\textsubscript{f}, \mu}[x_0]-\bar{x}_f\\
    \EX_{\vec{y}\textsubscript{f}, \mu}[z_0]-\bar{z}_f\\
    \EX_{\vec{y}\textsubscript{f}, \mu}[v_{x,0}]-\bar{v}_{x,f}\\
    \EX_{\vec{y}\textsubscript{f}, \mu}[v_{z,0}]-\bar{v}_{z,f}\\    
    \EX_{\vec{y}\textsubscript{f}, \mu}[C]-\bar{C}
    \end{cases}
\text{,}
\end{equation}
 where $\mu$ is the mass ratio parameter and $\bar{C}$ is the average Jacobi constant.
Since the final state and the dynamics are uncertain, the expected value of the initial state, which is found by backpropagating the final state, needs to be taken over both the reduced state and the mass ratio parameter. This is indicated by the use of the subscripts.  
Essentially, the idea is to start from a given final covariance, and backpropagate until the initial conditions on the Poincaré map, verifying whether solutions exist that maintain the periodicity condition, on average. Finding zeros of $\vec{G}$ implies finding initial conditions for the reduced state that ensure that its pdf maintains the same average periodic orbit and energy, while also maintaining a bounded covariance, with user-defined values. 
In our study, we focus on the dynamics of the CR3BP for distant retrograde orbits and northern Halo orbits, which we discuss in Sec.~\ref{sec:periodic_orbits_in_the_stochastic_system} and ~\ref{sec:stochastic_cont_northern_halo}. It is important to point out that the existence and uniqueness of these orbits are not guaranteed (since it is related to the existence and uniqueness of periodic solutions of the Fokker-Planck equation). Factors like the initial covariance values, the orbit of interest, and the dynamical system being studied highly affect the existence of these natural orbits.

\subsection{Approximating Moments Evolution on Poincarè Maps}
\label{sec:da_based_moments_evolution_on_poincare_maps}

An essential step in performing the discussed stochastic continuation procedure is to map initial uncertainties onto Poincarè maps. While this could be achieved via sample-based methods, stopping the integration at the surface of section, we propose an alternative procedure, based on the algebra of truncated polynomials.

Leveraging the algebra of truncated polynomials\footnote{the computer implementation of the algebra of truncated polynomials is often termed "differential algebra"~\cite{berz1989differential}.}~\cite{makino2006cosy, rasotto2016differential, Izzo2018}, we can first write future deviations (i.e., $\delta\vec{x}$) of the state after one period as a vector of Taylor polynomials that only depend on initial state and parameters deviations. It is important to note that this is only one of the possible methods to perform this (other options are the integration of high-order variational equations, also referred to as state transition tensors, or high-order automatic differentiation). In the case of Eq.~\eqref{eq:equations_of_motion}, we can write deviations of the solution of the ODEs after one period T as:
\begin{equation}
    \delta x^i_f= \mathcal{P}_{i}^k(\delta \vec{x}_0, \delta \mu, \delta T)+\mathcal{O}(k)
    \label{eq:taylor_expansion}
\text{,}
\end{equation}
where $\delta x^i_f$ is the $i$-th component of the state deviation at the final integration time, while $\mathcal{P}_{i}^k$ is the $k$-th order Taylor polynomial for the $i$-th component of the final state deviations that describes how deviations in the initial state, $\delta \vec{x}_0$, mass ratio parameter $\delta \mu$, and period $\delta T$, are mapped at future times. Furthermore, we indicate with $\mathcal{O}(k)$ all the terms of the series that have an order higher than $k$, which are ignored when constructing the Taylor series of order $k$. For constraining ourselves to the study of these solutions at a Poincaré section, we impose an additional constraint on the $i$th-component of the state. Without loss of generality, in the reminder of this paper and in the experiments shown in Sec.~\ref{sec:experiments}, the $y$-component of the state (i.e., $i=1$) is selected, hence:
\begin{equation}
    y_f-y_0=0
    \text{.}
    \label{eq:surface_of_section}
\end{equation}
 Then, we invert the map made by Eq.~\eqref{eq:surface_of_section} and \eqref{eq:taylor_expansion}, and we find the period correction that enables the satisfaction of Eq.~\eqref{eq:surface_of_section}. This will also be a Taylor polynomial:
\begin{equation}
    \delta T_{sos}=\delta T_{sos}(\delta \vec{x}_0, y_f-y_0=0, \delta \mu)\text{,}
\end{equation}
which we can then substitute in Eq.~\eqref{eq:taylor_expansion} to yield:
\begin{equation}
    \delta \vec{y}=\delta \vec{y}(\delta x^j_0, \delta \mu),\ \forall j\neq i 
    \text{,}
\end{equation}
also known as partial map inversion~\cite{armellin2010asteroid, fu2024high}.
Finally, we have reached a Taylor polynomial that describes how initial deviations in the reduced state and mass ratio parameter, are mapped into deviations of the reduced state, at the imposed surface of section. This means that we have a Taylor expansion of the state deviations at the Poincaré map of interest.
Using multi-index notation~\cite{reed1980methods} and grouping the deviations as $\delta\vec{p}=[x^j_0, \delta \mu]^T \ \forall j\neq i$, the vector of Taylor polynomials of order $k$ of the deviations of the state vector at the Poincaré map can be written as (neglecting the Taylor expansion terms higher than $k$):
\begin{equation}
    \delta\vec{y}\approx\vec{\mathcal{P}}^k(\delta \vec{p})=\sum_{|\alpha|=0}^k \dfrac{1}{\alpha!}(\partial^\alpha x)\bigg|_{(\vec{x}_0,\mu,T)}\delta \vec{p}^\alpha\text{,}
    \label{eq:taylor_polynomials_multi_index}
\end{equation}
where $k$ is the order of the Taylor polynomial expansion, and $|\alpha|=\sum_{j=0}^n\alpha_j$ must be taken over all possible combinations of $\alpha_j\in \mathbb{N}$. 
By taking the expected value of the vector of deviations in Eq.~\eqref{eq:taylor_polynomials_multi_index}, noting that on the right-hand side, the only terms that are affected by the expectation operator are the deviation, and due to the linearity of this operator, we can write~\cite{park2006nonlinear, park2007nonlinear, valli2013nonlinear}:
\begin{equation}
    \EX[\delta\vec{y}]\approx \sum_{|\alpha|=0}^k \dfrac{1}{\alpha!}(\partial^\alpha x)\bigg|_{(\vec{x}_0,\mu,T)}\EX[\delta \vec{p}^\alpha]\text{.}
    \label{eq:expected_value_poincare_map_deviations}
\end{equation}
By expressing the state deviations via Taylor polynomials of the parameters and initial state deviations, we have therefore reduced the computation of the expectation of the state deviations, to the computation of the expectation of the initial state and parameter deviations. Then, instead of assuming Gaussianity and applying Isserlis' theorem to simplify the computation of these terms~\cite{isserlis1918formula}, we leverage moment generating functions, with the only requirement that for this method to apply, the probability density function needs to admit a moment generating function.
These functions are powerful for this application since it is known that the $k$-th moment of the probability density function is connected to the moment generating function as:
\begin{equation}
    \EX[X_1^{k_1}\dots X_{n}^{k_n}]=\dfrac{\partial^k}{\partial t_1^{k_1}\dots \partial t_{n}^{k_n}}M_{\vec{X}}(\vec{t})\bigg|_{\vec{t}=\vec{0}}
    \text{,}
    \label{eq:moment_generating_function_and_moments}
\end{equation}
where $k_1,\dots, k_n$ are nonnegative integers, and $k=k_1+\dots+k_n$,
where $X_1,\dots, X_n$ are the $n$-components of a random variables vector $\vec{X}$. Hence, using Eq.~\eqref{eq:moment_generating_function_and_moments}, we can compute, symbolically, all the expectation terms on the right-hand side of Eq.~\eqref{eq:expected_value_poincare_map_deviations}: this can be done only once for each pdf of interest, offline and then be re-used and evaluated at the specific parameters that describe the pdf. For this purpose, we use the expression system of \textit{heyoka}'s software\footnote{\url{https://bluescarni.github.io/heyoka.py}, date of access: January 2024}.
For the case of a Gaussian, for instance, which we will use across Sec.~\ref{sec:experiments}, the moment generating function can be expressed as:
\begin{equation}
    M_{\pmb{X}}(\pmb{t})=\textrm{exp}(\vec{\mu}^T\pmb{t}+\dfrac{1}{2}\pmb{t}^T\Sigma \pmb{t})
    \text{,}
    \label{eq:moment_generating_function}
\end{equation}
where $\vec{\mu}\in \mathbb{R}^n$ and $\Sigma\in\mathbb{R}^{n\times n}$ are the mean vector and covariance matrix of an $n$-dimensional random vector, respectively, while $\pmb{t}\in \mathbb{R}^n$. 
Hence, we have obtained a method to directly compute the moments of the propagated state at the Poincaré map, at the expense of having to compute only the state transition tensor elements of the flow after one period, which we perform via \textit{pyaudi}\footnote{\url{https://darioizzo.github.io/audi/}, date of access: January 2024.}.

\section{Numerical Simulations}
\label{sec:experiments}
\subsection{Poincaré Map Reconstruction}
\begin{figure}[hbt!]
\centering
\includegraphics[width=0.4\textwidth]{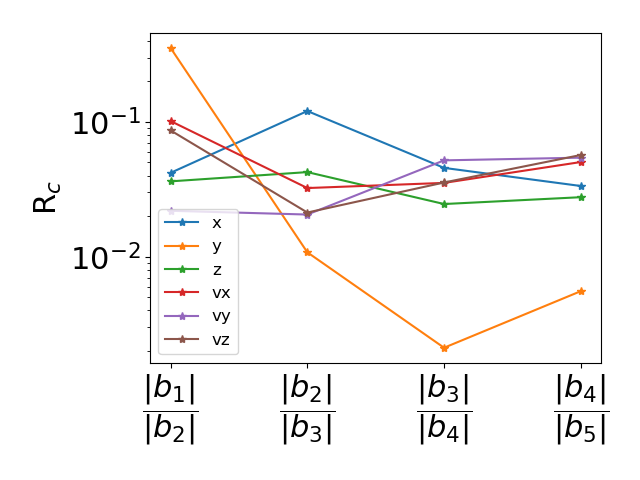}
\caption{Ratio test analysis.}
\label{fig:ratio_test}
\end{figure}

\begin{figure*}[hbt!]
\centering
\begin{subfigure}{0.4\textwidth}
  \centering
  \includegraphics[width=\linewidth]{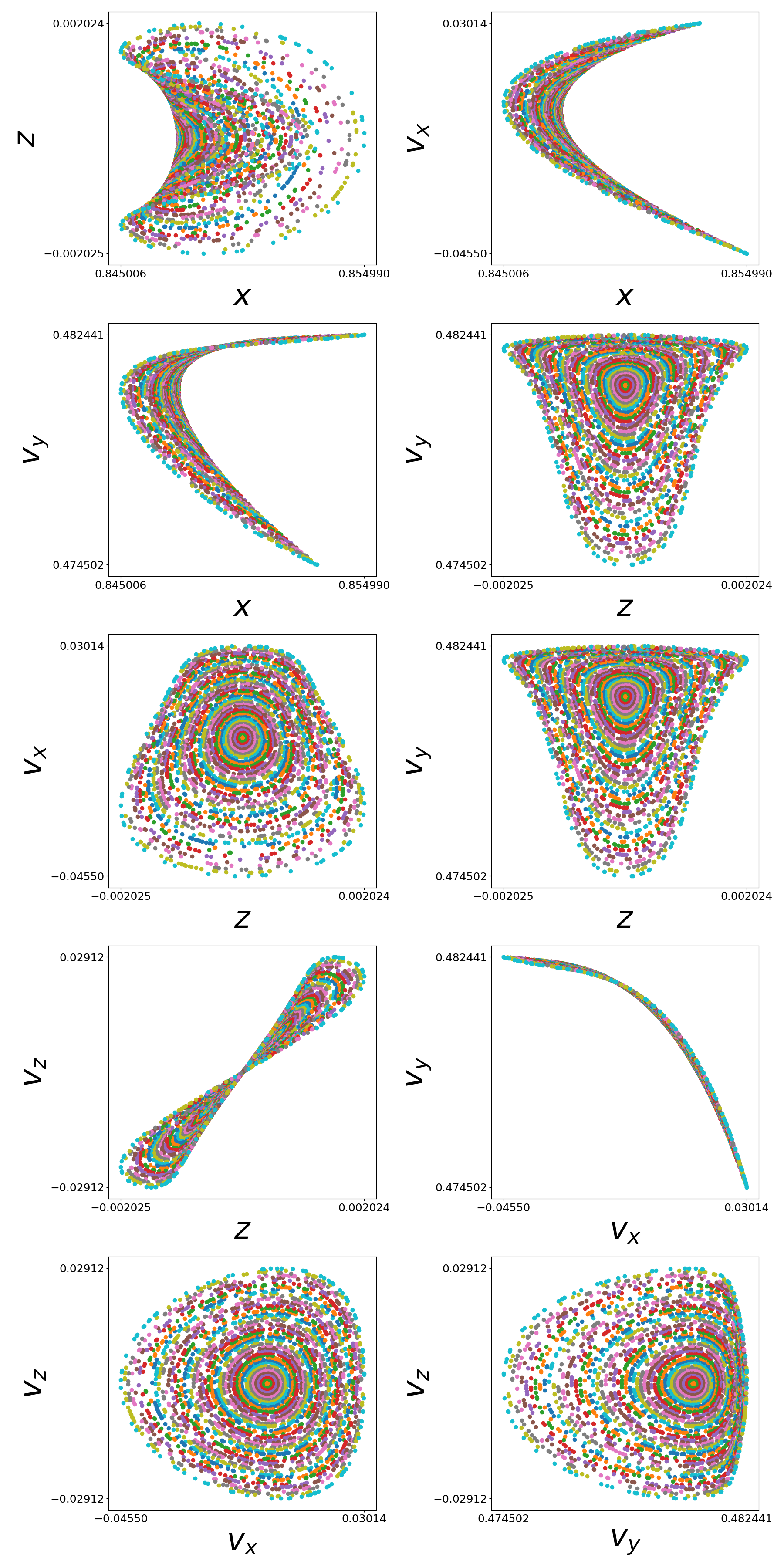}
  \caption{Ground Truth}
  \label{fig:ground_truth_poincare}
\end{subfigure}
\begin{subfigure}{0.4\textwidth}
  \centering
  \includegraphics[width=\linewidth]{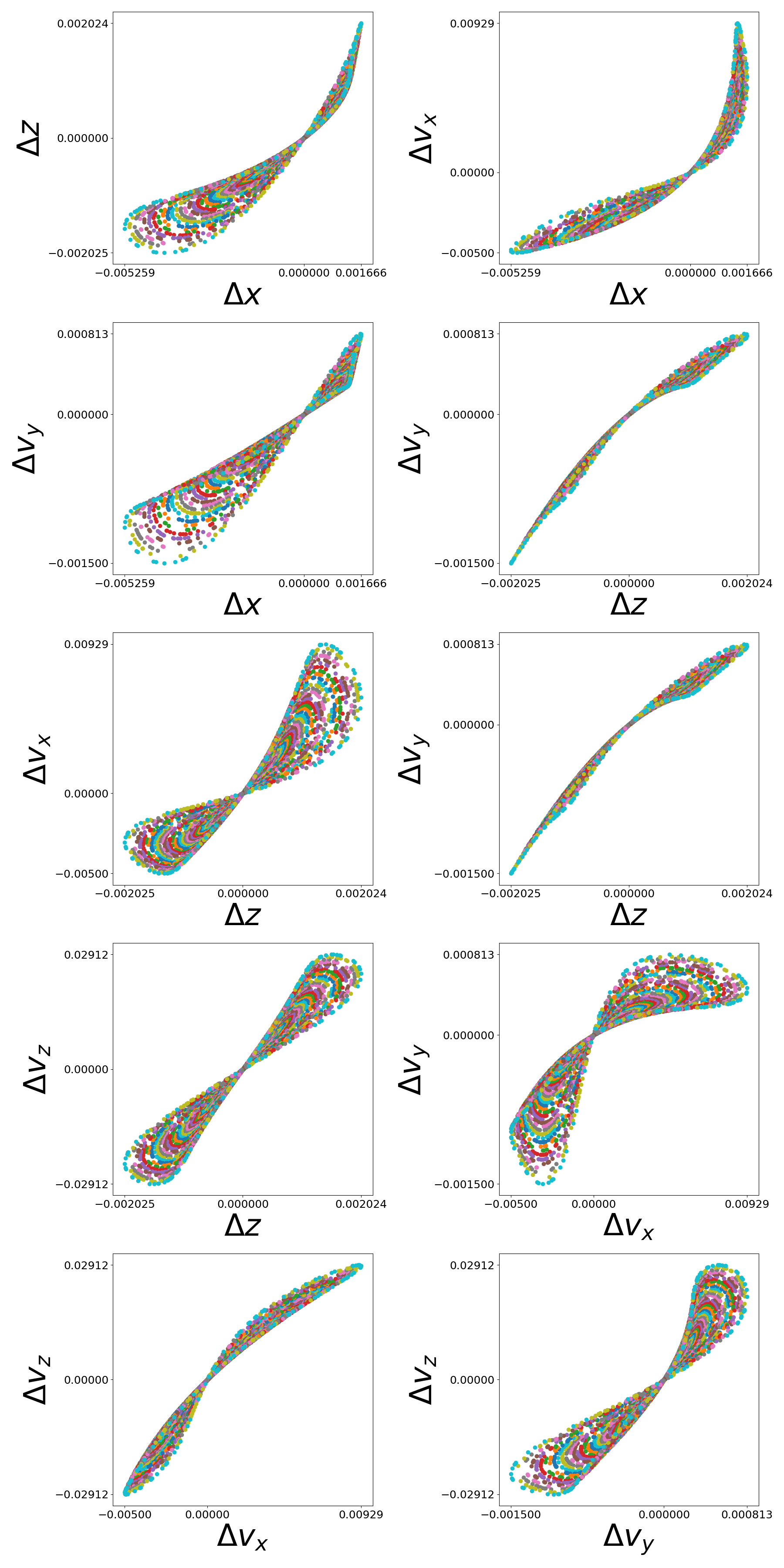}
  \caption{Errors.}
  \label{fig:errors_da_poincare}
\end{subfigure}
\caption{Ground truth (\textit{left}) and errors (\textit{right}) in the Poincaré map construction. Errors are defined as ground truth minus DA approximation.}
\label{fig:da_approximation_poincare}
\end{figure*}

As an initial validation of our method, we compare how well the Taylor expansions can reconstruct the Poincaré map in the CR3BP. To do this, we focus on distant retrograde orbits, and we constrain the analysis at the surface of section generated via successive crossings of the spacecraft on the plane identified by $y_f-y_0=0$, where the state is written as $\vec{x}=[x,y,z,v_x,v_y,v_z]^T$. Furthermore, we impose perturbations on the state w.r.t. the nominal case that range from a five-dimensional ball of radius $10^{-5}$ to $1.5\times 10^{-2}$ in normalized units (corresponding to maximum distances of thousands of kilometers).
As an initial analysis of the convergence radius, the ratio test (i.e., D'Alembert's criterion) can also be used. This computes the convergence radius as:

\begin{equation}
    \epsilon= \lim_{k\rightarrow \infty}\dfrac{|b_k|}{|b_{k+1}|}\text{,}
\end{equation}

where $b_k$ are the matrices resulting from unfolding the tensor of the $k$-th Taylor term~\cite{izzo2020stability}. In practical cases, the series must be truncated at a given finite $k$, and the radius is therefore approximated without all coefficients. For an analysis of the behavior of this radius as a function of the order of the coefficients, in Fig.~\ref{fig:ratio_test}, we show an example of how the radius of convergence progresses, for all the different state components, for a given member of the DRO family, with an initial state: $\vec{x}_{po}=[8.47361113\times10^{-1},0,0,8.08932591\times10^{-15}, 4.80694267\times10^{-1},0]^T$, mass ratio parameter: $\mu= 0.01215058$, and period: $T=2.35248$. As we can observe, the convergence radius seems to stabilize around values of about $5\times 10^{-3}$ at worst, for already order 4. 
To confirm this, in Fig.~\ref{fig:da_approximation_poincare}, we also display the actual behavior of the Taylor approximation, compared to the ground truth solution (found by propagating the Poincaré map numerically). As it is observed qualitatively and also quantified numerically, the errors start to grow 'uncontrollably', for perturbations of around $5\times 10^{-3}$, confirming the relatively good approximation of the ratio test estimate for the case of interest.
\subsection{Uncertainty Propagation}
\begin{figure}[htb!]
    \centering
    \begin{subfigure}[b]{0.5\linewidth}
        \centering
        \includegraphics[width=\linewidth]{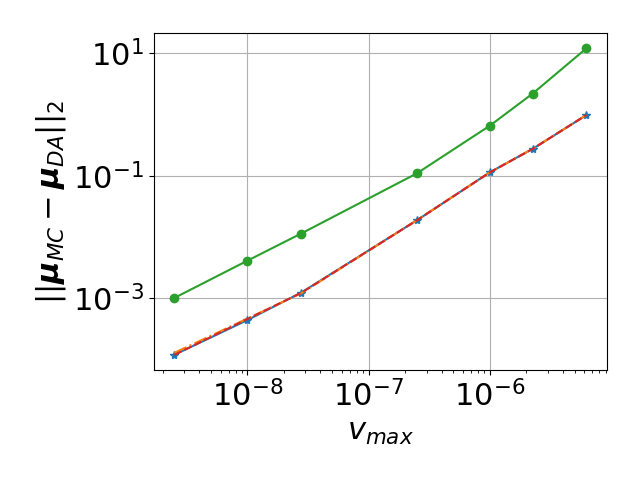}
    \end{subfigure}
    \hfill
    \begin{subfigure}[b]{0.5\linewidth}
        \centering
        \includegraphics[width=\linewidth]{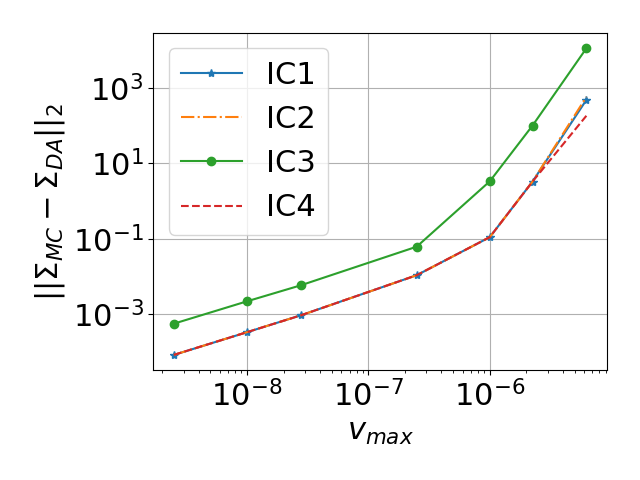}
    \end{subfigure}
    \hfill
    \begin{subfigure}[b]{0.5\linewidth}
        \centering
        \includegraphics[width=\linewidth]{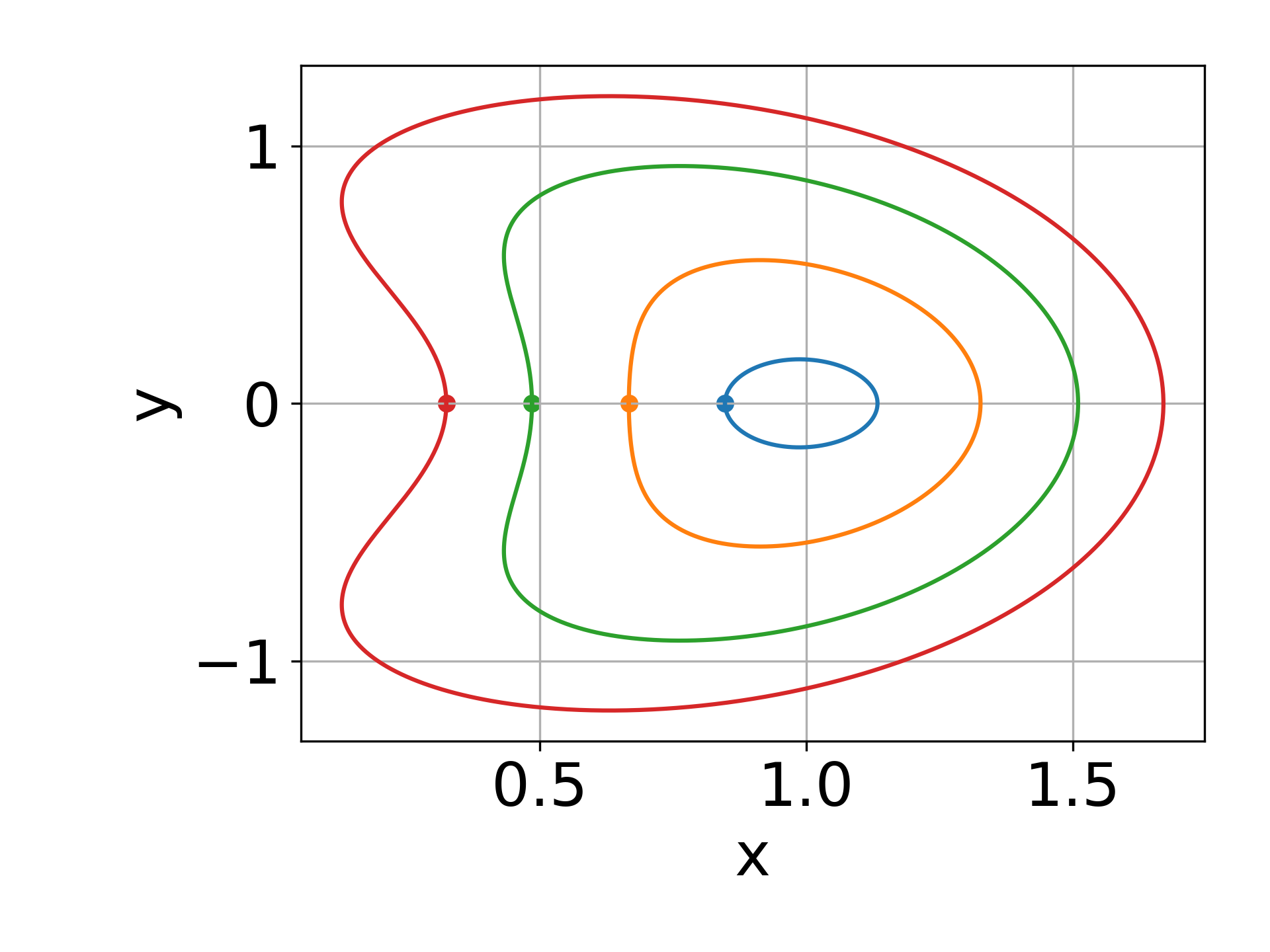}
    \end{subfigure}
    \caption{\textit{top}: 1st-moment error in L2-norm between DA and MC methods, as a function of maximum covariance matrix perturbation; \textit{center}: 2nd-moment error in Frobenius norm between DA and MC methods, as a function of maximum covariance matrix perturbation; \textit{bottom}: initial conditions selected from the DRO families.}
    \label{fig:error_up_da_vs_mc}
\end{figure}

As a second step, we analyze how well the moments of the state at the Poincaré map can be reconstructed, using the approach discussed in Sec.~\ref{sec:da_based_moments_evolution_on_poincare_maps}. To do this, we select four initial conditions from the distant retrograde orbits family, leveraging the planar circular restricted three-body problem, shown on the bottom part of Fig.~\ref{fig:error_up_da_vs_mc}, and we assume that the initial conditions (marked in the plot with a dot) and mass ratio parameter are affected by uncertainties of different magnitudes. These range from a standard deviation of $5\times 10^{-6}$ to $2.5\times10^{-4}$ in the mass ratio parameter, and a diagonal Cholesky matrix with terms from $\sigma_{xx}=2.5\times10^{-3}$, $\sigma_{v_xv_x}=2\times10^{-4}$, $\sigma_{v_yv_y}=2\times10^{-4}$ to $\sigma_{xx}=0.125$, $\sigma_{v_xv_x}=0.01$, $\sigma_{v_yv_y}=0.01$ in the initial state. In the top of Fig.~\ref{fig:error_up_da_vs_mc}, we display how the error in the L2-norm between the DA-based and Monte Carlo-derived mean progresses, as a function of the magnitude of maximum stretch in the covariance matrix, while in the central plot of Fig.~\ref{fig:error_up_da_vs_mc}, we do the same plot, but for the difference in propagated covariance matrices. 
As we observe from the deviations of the DA-based mean and covariance matrix approximations, w.r.t. those found via Monte Carlo sampling (with one million samples), the accuracy of the approximation seems to degrade substantially for larger covariances, and it also seems to behave in a similar way for periodic orbits of the same family, although there can be some exceptions (like IC3). From this preliminary analysis, it is clear that before concluding that the DA-based approximation can fulfill the required accuracy level, specific families and periodic orbit members as well as types and order of magnitude of uncertainties have to be studied in detail. In the DRO planar family case, we observe that for maximum initial perturbations of the level of about  $2,000-3,000$ km in position, which are more than enough for practical applications, the DA-based method seems to very well capture the underlying behavior of the moments of the distribution.

\subsection{Stochastic Continuation on DROs}
\label{sec:stochastic_cont_dro}

\begin{figure}[htb!]
    \centering
    \begin{subfigure}[b]{0.49\linewidth}
        \centering
        \includegraphics[width=\linewidth]{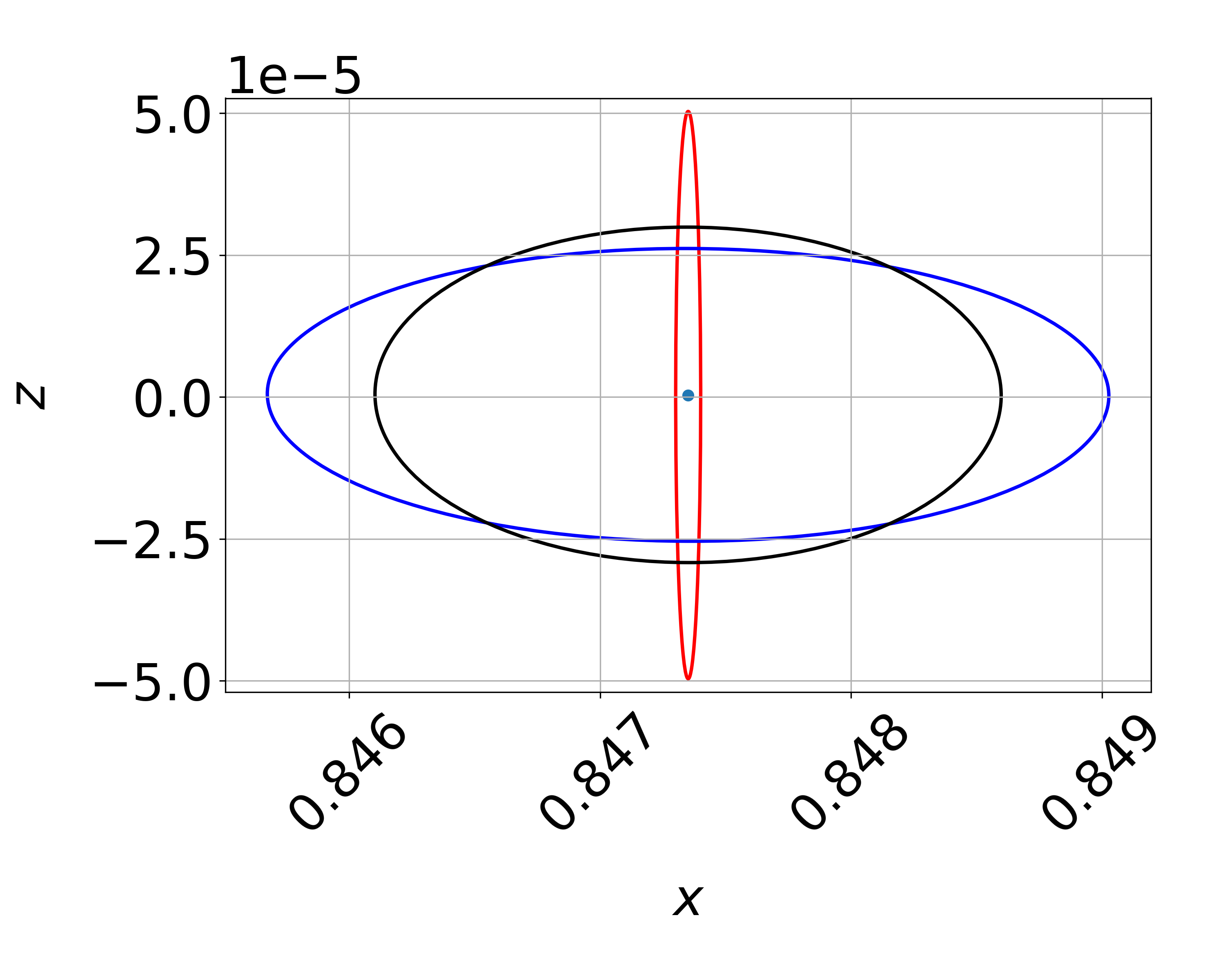}
    \end{subfigure}
    \hfill
    \begin{subfigure}[b]{0.49\linewidth}
        \centering
        \includegraphics[width=\linewidth]{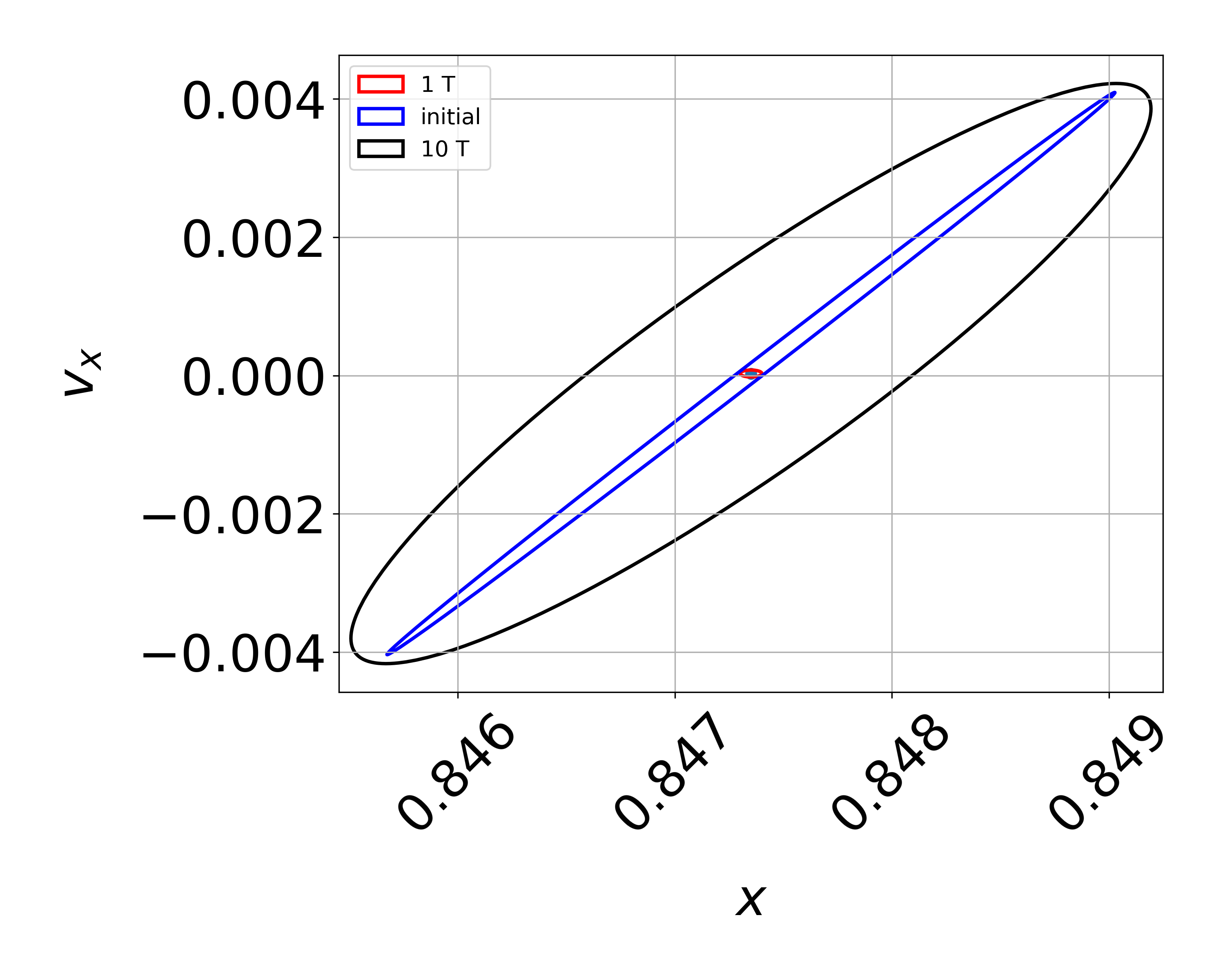}
    \end{subfigure}
    \hfill
    \begin{subfigure}[b]{0.49\linewidth}
        \centering
        \includegraphics[width=\linewidth]{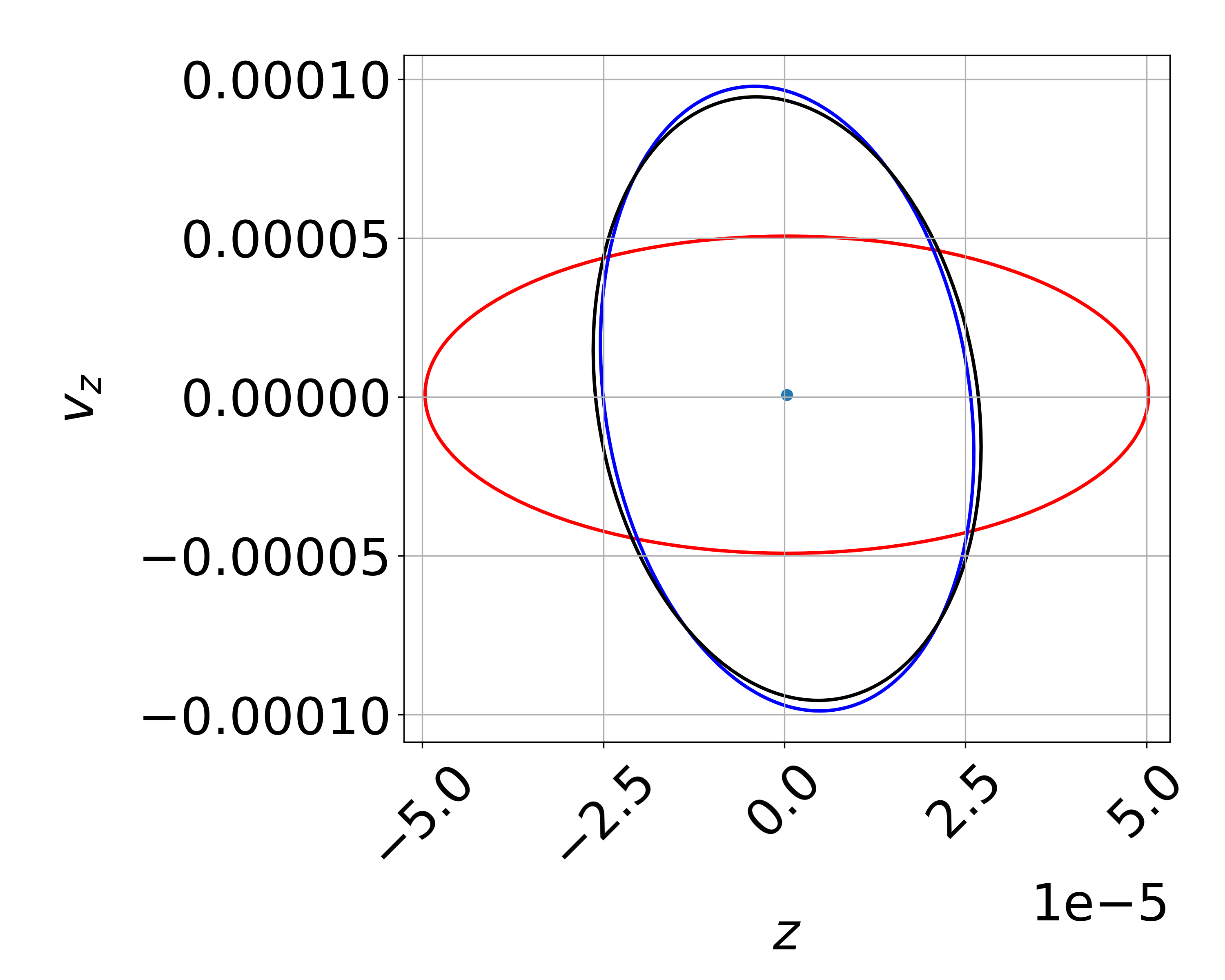}
    \end{subfigure}
    \caption{Three Poincaré map projections for the converged solution in the DRO family case.}
    \label{fig:ellipsoids_dro_projections}
\end{figure}

Leveraging the setup discussed in Sec.~\ref{sec:periodic_orbits_in_the_stochastic_system}, we display an example of differentially correcting a deterministic three-dimensional DRO family member. In particular, we select a DRO deterministic orbit in the Earth-Moon system, with the initial state (in normalized units): $\vec{x}_0=[8.47361113\times10^{-1},0,0,8.08932591\times10^{-15},4.80694267\times10^{-1},0]^T$. We decided to start with the DRO family because they are particularly stable, showcasing large stability regions surrounding their nominal orbit. We impose to find solutions that after one period, fulfill a covariance matrix in the reduced state of $\Sigma_{\vec{y}_f}=\textrm{diag}([2.5,2.5,2.5, 2.5, 2.5]^T)\times10^{-9}$. The mass ratio parameter was also assumed to be a random variable, sampled from a Gaussian distribution of standard deviation: $\sigma_{\mu}=10^{-4}$. The stochastic continuation method converged to an average initial periodic orbit with the following state: $\bar{\vec{x}}_0=[8.47350680\times10^{-1}, 0, 2.15211495\times10^{-7}, 2.95109959\times10^{-5}, 4.80691639\times10^{-1}, 2.90325842\times10^{-7} ]$. 
In Fig.~\ref{fig:ellipsoids_dro_projections}, we display some of the projections of the found bounded region. As we see, the method is able to find a region where the motion stays within the requested covariance ellipsoid, after one orbit. Furthermore, we also show (in black) the behavior of the same trajectories after 10 orbits: showing that these will not tend to diverge, but will maintain a bounded behavior w.r.t. the nominal orbit. This happens due to the overall stability of these orbits.
\subsection{Stochastic Continuation on Northern Halo}
\label{sec:stochastic_cont_northern_halo}
As a final step, we also investigated the use of the technique for the northern Halo family in the Earth-Moon system. In particular, we used the technique starting from a Halo orbit with only center manifolds, similar to the case of the DRO. For this, the periodic orbit with the following initial reduced state was chosen as initial guess: $\vec{y}_0=[8.76354295\times10^{-1},1.91924044\times10^{-1},-4.99033766\times10^{-14},2.30070207\times10^{-1},1.48089385\times10^{-13}]^T$. The final covariance matrix of the reduced state and the mass parameter standard deviation were imposed to be identical to the one used for the DRO case, in Sec.~\ref{sec:stochastic_cont_dro}. 

\begin{figure}[htb!]
    \centering
    \begin{subfigure}[b]{0.49\linewidth}
        \centering
        \includegraphics[width=\linewidth]{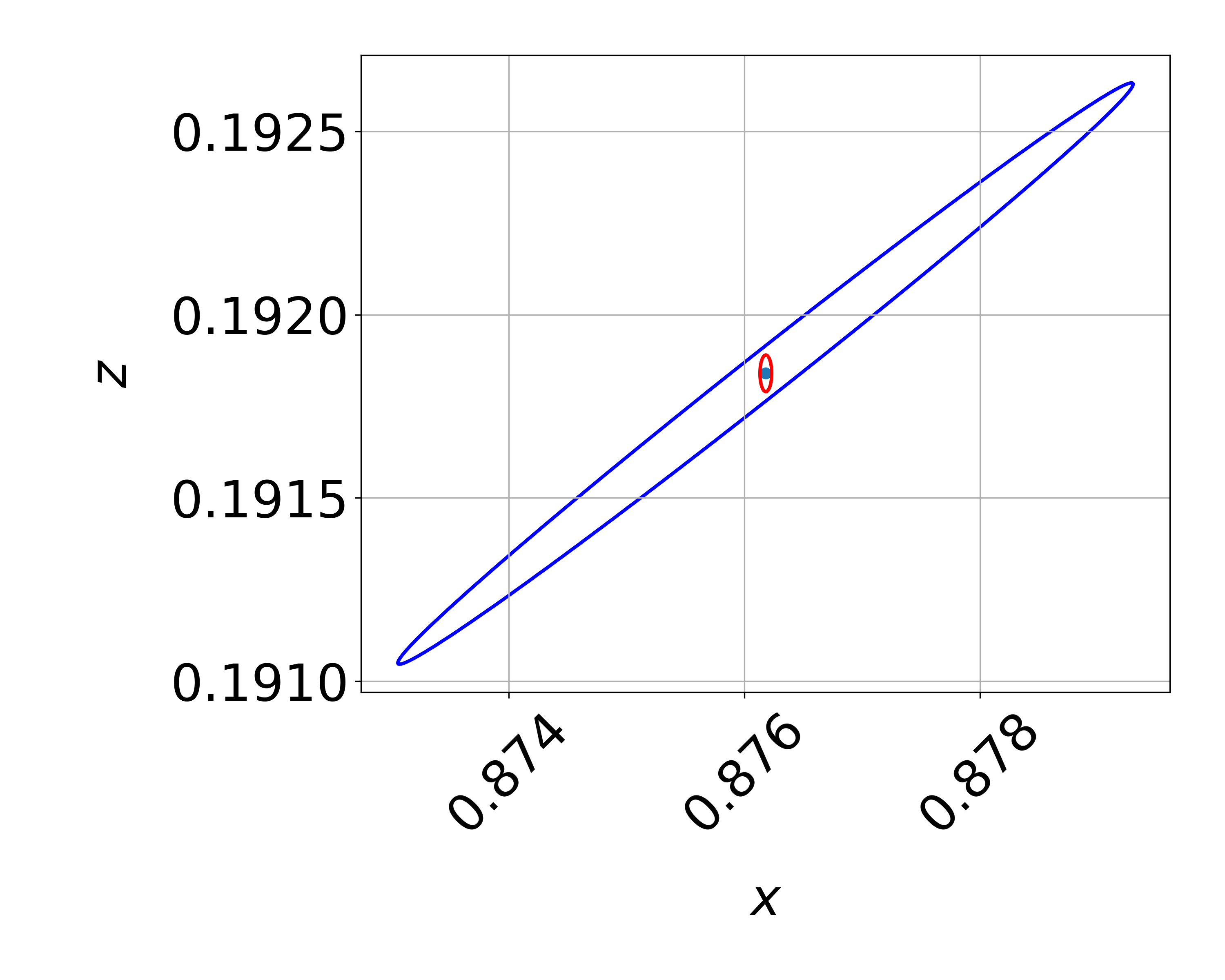}
    \end{subfigure}
    \hfill
    \begin{subfigure}[b]{0.49\linewidth}
        \centering
        \includegraphics[width=\linewidth]{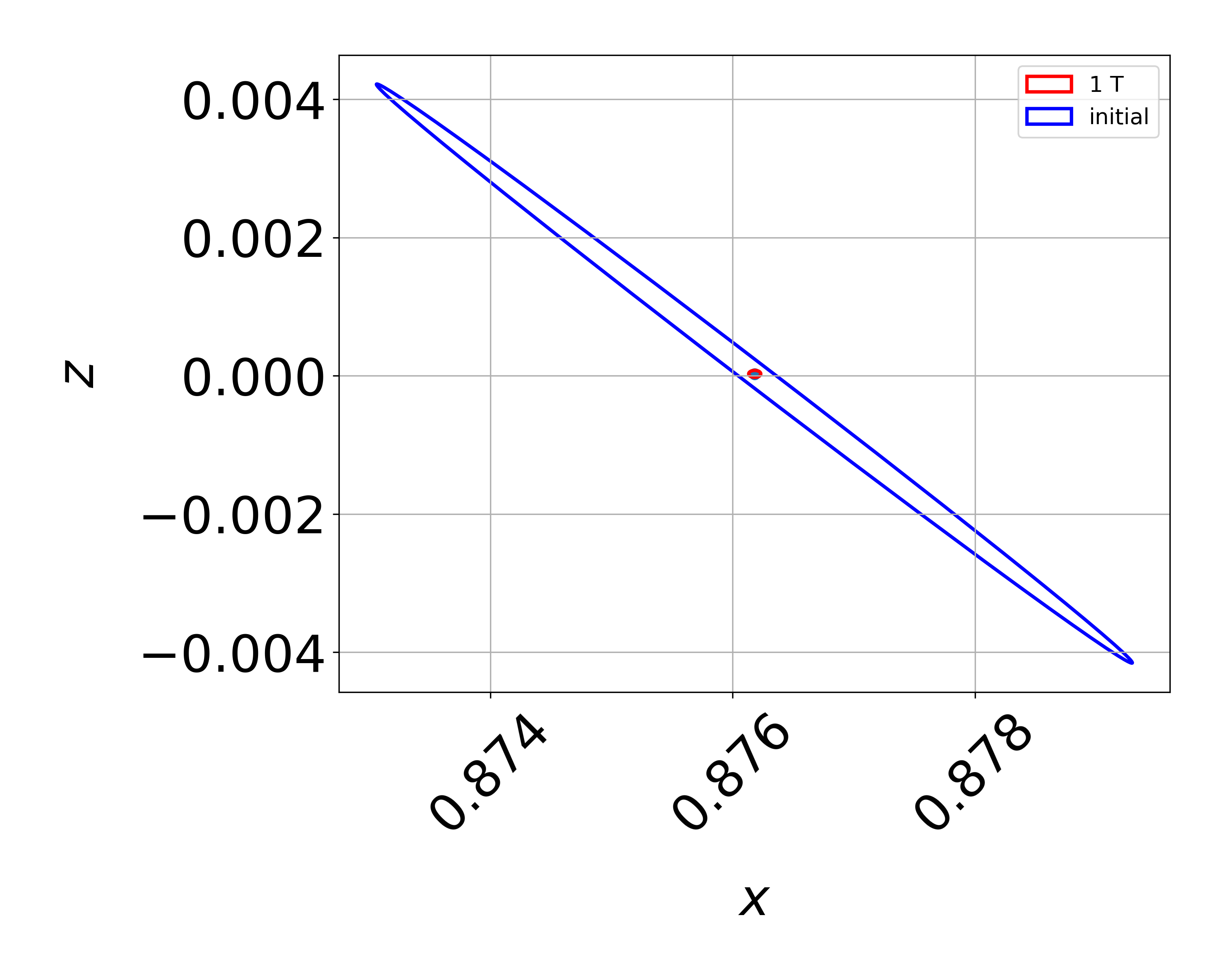}
    \end{subfigure}
    \hfill
    \begin{subfigure}[b]{0.49\linewidth}
        \centering
        \includegraphics[width=\linewidth]{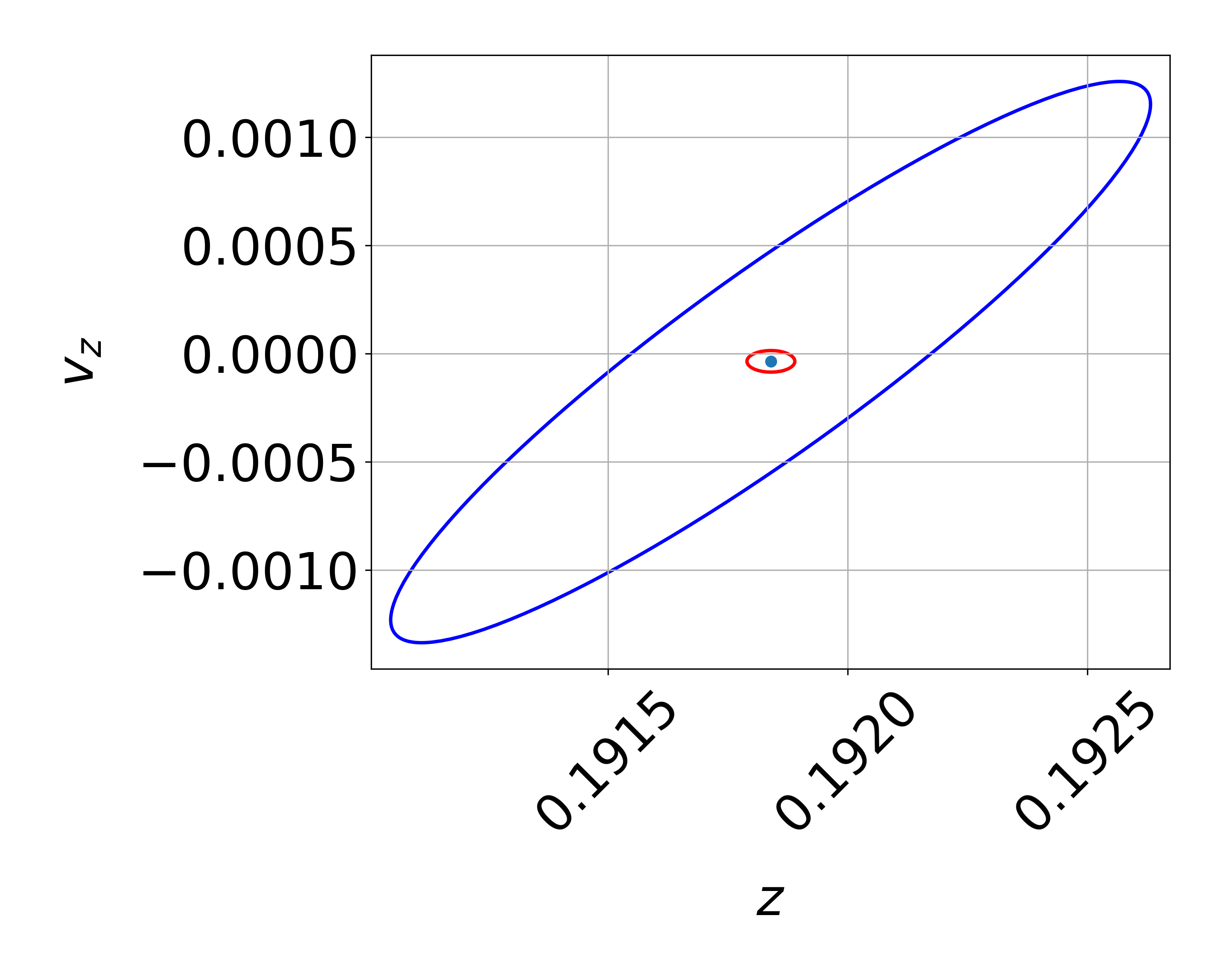}
    \end{subfigure}
    \caption{Three Poincaré map projections for the converged solution in the northern Halo family case.}
    \label{fig:ellipsoids_northern_halo_projections}
\end{figure}

In this case, the method was able to converge to solutions that maintain a bounded covariance after one period, within the user-defined thresholds. However, the technique converged to a solution that is surrounded by stable and unstable manifolds, and trajectories around this average periodic orbit are only temporarily captured (at one period) within a region defined by $\Sigma_{\vec{y}_f}$, before then diverging after one period. The converged average periodic orbit has the following initial reduced state: $\bar{\vec{y}}_0=[8.76181969\times10^{-1},1.91839767\times10^{-1},2.72156355\times10^{-5},2.30367787\times10^{-1},-3.63731138\times10^{-5}]^T$, which is substantially different than its deterministic counterpart.
In Fig.~\ref{fig:ellipsoids_northern_halo_projections}, we show some projections of such converged solutions. 
After one orbit the final state covariance on the Poincaré map reaches the required values while maintaining the average periodicity condition. However, by integrating the solution for multiple orbits, the trajectories quickly diverge along the unstable manifold (reaching covariances three orders of magnitude larger, after only ten orbits). This aspect highlights that the technique can help in finding natural regions of bounded motion, in case they exist, but their existence, uniqueness, and asymptotic behavior highly depends on the problem specifics. For a more general formulation that maintains a bounded region for any dynamical environment and uncertainties, a control strategy must be accounted for.

\section{Conclusions}

In this work, we have discussed a technique to continue periodic orbits in the CRTBP when errors in the initial conditions of the spacecraft and mass ration parameter between the primaries are taken into account. In particular, we have focused on the formulation of the stochastic continuation procedure for the study of some members of three-dimensional distant retrograde orbits and northern Halo orbits, and on the formulation and integration of a DA-based technique to prevent the computational burden of propagating samples at the Poincaré map using Monte Carlo sampling. 

First, as it is shown via numerical experiments for the DRO family, the DA-based method successfully manages to both reconstruct the Poincaré map and approximate the expected values on the map, for relatively large convergence radii. 
Second, the proposed stochastic continuation formulation successfully manages to find regions of bounded motion around the periodic orbit, while also maintaining the periodicity condition, on average. In this way, the technique might serve as a useful tool for preliminary mission design in uncertain environments.

While the stochastic continuation approach is general in nature, the existence and uniqueness of solutions closely influences the capability of the method to find solutions, as was proven by the northern family Halo example discussed in this paper. Hence, without any form of control, finding bounded solutions within user-defined tolerances might turn out to be unfeasible. For this reason, control techniques will be explored to make the technique more generally applicable for any mission scenario, provided that the necessary corrections are applied.

\section*{Acknowledgments}

This work has been funded by the Open Space Innovation Platform (OSIP) of the European Space Agency.

\bibliographystyle{ISSFD_v01}
\bibliography{references}

\end{document}